\documentclass{PoS}

\newcommand{\MSSM}{\textmd{\textsc{mssm}}}
\newcommand{\Mh}{\ensuremath{M_h}}
\newcommand{\SUSY}{{\sc susy}}
\newcommand{\QCD}{{\sc qcd}}
\newcommand{\SUSYQCD}{\SUSY-\QCD}
\newcommand{\LHC}{{\sc lhc}}
\newcommand{\DRBAR}{\ensuremath{\overline{\textmd{\textsc{dr}}}}}
\newcommand{\SPS}[1]{\textsc{sps#1}}

\newcommand{\Mt}{\ensuremath{m_t}}

\usepackage{psfrag}
\usepackage{amsmath}
\usepackage{amssymb}

\usepackage{cite}

\title{\vskip-3\baselineskip%
  {%
    \normalsize\textrm{SFB/CPP-10-70}
  }%
  \vskip-2\baselineskip%
  \vskip3\baselineskip%
Three-loop Corrections to the Mass of the Light Higgs Boson in the MSSM}

\ShortTitle{Three-loop Corrections to the Mass of the Light Higgs Boson in the MSSM}

\author{\speaker{Philipp KANT}\\%
        TTP Karlsruhe\\
        E-mail: \email{philipp.kant@kit.edu}}

      \abstract{The Minimal Supersymmetric extension of the Standard
        Model (\MSSM) predicts the existence of a light neutral Higgs
        boson. Once found at the \LHC, its mass will immediately become
        a precision observable. The theoretical value of the Higgs mass
        \Mh{} is subject to large radiative corrections. Due to the
        large top Yukawa coupling, loops of top quarks and their
        superpartners provide the dominant contribution to the radiative
        corrections.  

        We present a calculation of the \SUSYQCD{} corrections to these
        diagrams, up to the three-loop order. We find that our
        three-loop results can be in the range of one GeV, and are thus
        relevant when compared with the expected experimental accuracy
        at the LHC.  We also find a significantly reduced dependency on
        the renormalisation prescription, thus decreasing the
        theoretical uncertainty of the prediction of \Mh.}

\FullConference{European Physical Society Europhysics Conference on High Energy Physics\\
		 July 16-22, 2009\\
		 Krakow, Poland}

\begin{document}

\section{Introduction}

An important feature of the minimal supersymmetric extension of the
Standard Model (\MSSM) is the existence of a light neutral Higgs boson.
The \MSSM{} Higgs sector is a Two-Higgs Doublet Model, the parameters of
which are related to the gauge couplings through Supersymmetry.  These
relations reduce the number of new (in comparison with the Standard
Model) parameters to two, which are usually chosen to be the ratio
$v_2/v_1=\tan\beta$ of the vacuum expectation values of the two Higgs
doublets and the mass $M_A$ of the pseudoscalar Higgs.  Once these are
fixed, the mass \Mh{} of the light neutral Higgs boson is not a free
parameter, but a calculable prediction of the theory.
At the tree level, there is an upper limit
$\Mh \leq M_Z$.  At higher orders, this bound gets shifted up by
radiative corrections to the Higgs self-energy, which depend on the
spectrum of the superpartner masses.

The one- and two-loop corrections to \Mh{} have been extensively studied
in the literature (for reviews, see e.g.
Refs.~\cite{Heinemeyer:2004ms,Allanach:2004rh}).  The studies show that,
due to the large top-Yukawa coupling, the most sizeable corrections stem
from loops of top quarks and their superpartners in the Higgs propagator.
For a light Higgs boson, the Large Hadron Collider (\LHC) will be able
to measure its mass with an expected experimental accuracy of
$100-200\,$MeV~\cite{cmstdr}.  To take full advantage of the
experimental data, the theoretical prediction for \Mh{} has to match
this precision.  However, based mostly on the renormalization scale and
scheme dependence, the theoretical uncertainty on the prediction of
\Mh{} has been estimated to
$3-5\,$GeV~\cite{Degrassi:2002fi,Allanach:2004rh}.

While there have been many publications tackling the one- and two-loop
corrections to \Mh, there are only two calculations available that go to
the third order of perturbation theory.  In~\cite{Martin:2007pg},
Renormalisation Group methods have been used to calculate the leading-
and next-to leading term in $\ln(M_{SUSY}/M_t)$, where $M_{SUSY}$ is the
typical scale of \SUSY{} particle masses.  Motivated by the fact that
the corrections from top and stop loops dominate the overall
corrections, the three-loop \SUSYQCD{} corrections to these diagrams
have been computed in~\cite{Harlander:2008ju}.  Because current methods
are not sufficient to solve three-loop multiscale integrals exactly, the
calculation in~\cite{Harlander:2008ju} assumed a strong hierarchy among
the superparticle masses and performed nested asymptotic
expansions~\cite{Smirnov:2002pj} to reduce the problem to single-scale
integrals.  While~\cite{Harlander:2008ju} considered two rather simple
mass hierarchies, it is the aim of this talk to report on recent
progress on computing more involved scenarios and discuss which are
important for phenomenological studies.

\section{Outline of the Calculation}

We calculate the corrections to \Mh{} by evaluating virtual corrections
to the Higgs propagator.  As in~\cite{Harlander:2008ju}, we restrict
ourselves to diagrams where the Higgs couples to top quarks or their
superpartners, including \SUSYQCD{} corrections up to Order
$\mathcal{O}(\alpha_s^2)$.  This leaves us with the following virtual
particles: top quarks $t$ and their superpartners, the stops
$\tilde{t}_{1/2}$, the gluons $g$ and gluinos $\tilde{g}$, as well as
the light quarks $q$ and squarks $\tilde{q}$, which enter at the
three-loop level.

One can perform the calculation assuming different hierarchies amongst
the superpartner masses.  To estimate the error introduced by expanding
around these hierarchies, we systematically compare, at the two-loop
level, with the exact result which is given in~\cite{Degrassi:2001yf} in
very compact form.  
A detailed study of this comparison, which shows
that the relative error can be brought below $5\%$ for the \SPS{}
benchmark scenarios from~\cite{Allanach:2002nj}, will be
presented in a forthcoming publication~\cite{Kant:2010tf}.

\section{Renormalisation Prescription Dependence}

\begin{figure}
  {
    \input{figs/h3os-psfrag}
    \includegraphics[width=.5\textwidth]{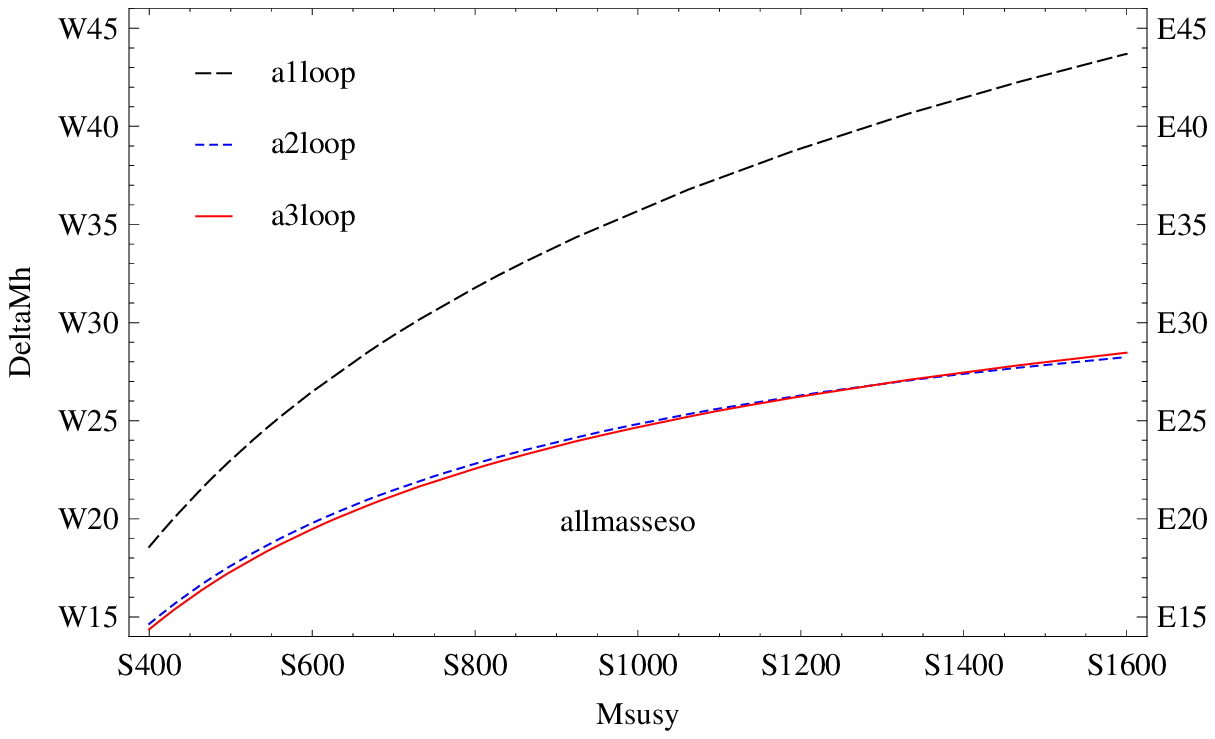}
  }
  {
    \input{figs/h3mixed-psfrag}
    \includegraphics[width=.5\textwidth]{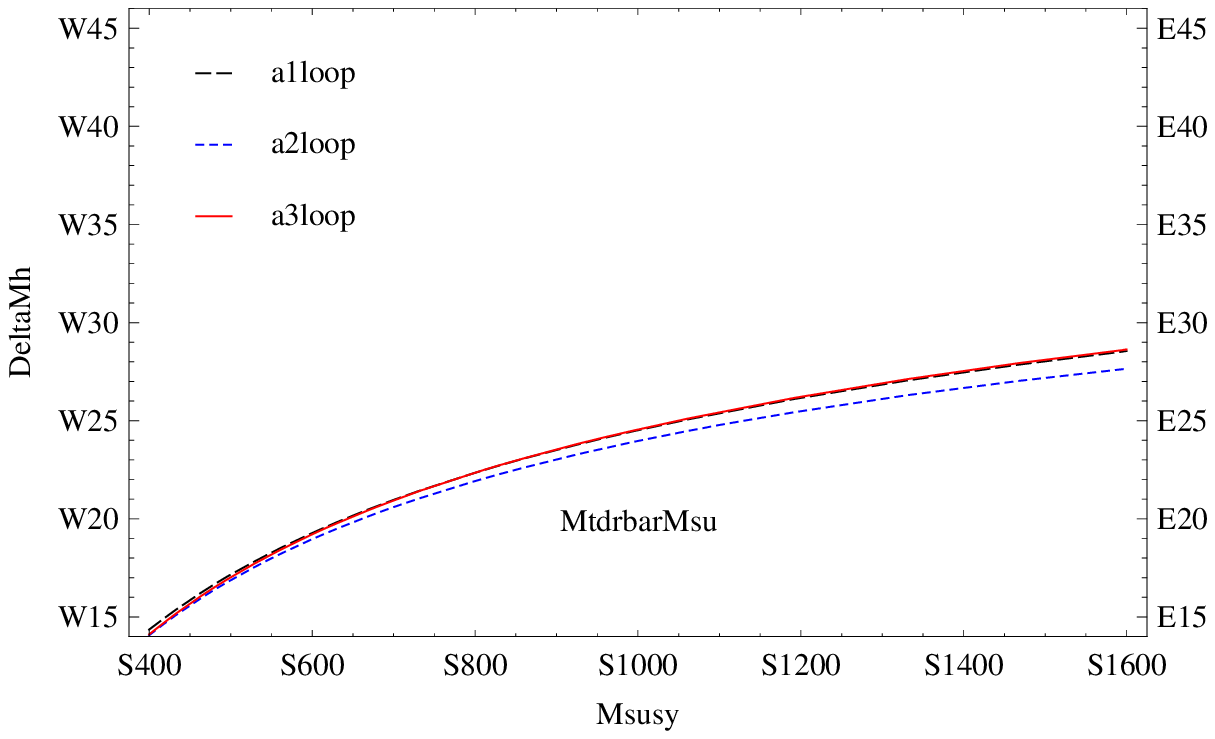}
  }
  \caption{Renormalisation scheme dependence of the corrections $\Delta
    \Mh = \Mh - \Mh^{tree}$.  This figure assumes a degenerate mass
    spectrum of the \SUSY{} particles.  In the left panel, all masses
    are renormalised on-shell, while in the right panel the \DRBAR{}
    scheme is used for the top mass.  The masses of the \SUSY{}
    particles are renormalised on-shell in both cases, to ensure that
    $\Delta \Mh$ is plotted over the same parameter.  Choosing the
    \DRBAR{} scheme also for the superpartner masses has a small effect
    on the Higgs mass.}
\label{fig:osdr}
\end{figure}

We use Dimensional Reduction~\cite{Siegel:1979wq} in order not to spoil
Supersymmetry through the regularisation.  This leaves us with the
choice between using on-shell renormalisation conditions and using
minimal subtraction, i.e. the \DRBAR{} scheme.  In order to make a
justified choice for one scheme over the other, we analysed the
magnitude of the corrections of different loop orders in both schemes
(see Fig.~\ref{fig:osdr}).

The figure demonstrates three points: {\it i)}, the large one-loop
corrections to \Mh{} are an artefact of using the on-shell top mass.
The corrections of higher orders are significantly smaller in the
\DRBAR{} scheme, indicating a better behaviour of the perturbative
expansion, {\it ii)}, while there is a significant deviation between the
two schemes at the one-and two-loop level, at the third loop order the
discrepancy all but vanishes, indicating a stabilisation of the
theoretical prediction, and {\it iii)}, the three-loop corrections
amount to some hundred MeV.
After this analysis, we resolved to use the \DRBAR{} scheme for the
renormalisation of our parameters.  An exception to this is the mass of
the $\varepsilon$-scalar that appears in Dimensional Reduction, which we
renormalise on-shell and set it to zero.

\section{Results}

\begin{figure}
  \begin{center}
    {
      \input{figs/sps2plot-psfrag.tex}
      \includegraphics[width=.8\textwidth]{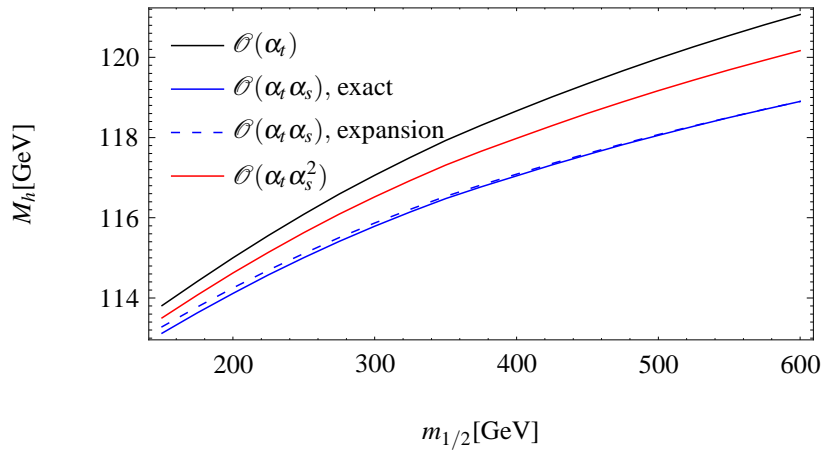}
    }
  \end{center}
  \caption{Higgs mass for the benchmark line \SPS{2}.  The red line
    includes our three-loop corrections, as well as all the corrections
    of one- and two-loop order that are implemented in {\sc FeynHiggs}.}
  \label{fig:sps2}
\end{figure}

As an example for the numerical impact of our calculation we show
results for the \SPS{} benchmark line \SPS{2}. Fig~\ref{fig:sps2}
shows \Mh{} including corrections of one-, two- and three-loop order
in black, blue and red, respectively.  To get the best prediction
possible, we extract all available one- and two-loop corrections from
the program {\sc FeynHiggs}~\cite{Frank:2006yh, Degrassi:2002fi,%
  Heinemeyer:1998np, Heinemeyer:1998yj}.  In particular, we use the
exact two-loop result from~\cite{Degrassi:2001yf} for the
$\mathcal{O}(\alpha_t\alpha_s)$ corrections.  It is notable that the
three-loop corrections amount to about one GeV for large $m_{1/2}$.

\end{document}